\documentstyle[twoside,fleqn,espcrc2,epsf]{article}

\let\ga=\gamma

\let\De=\Delta

\let\La=\Lambda

\let\si=\sigma

\let\om=\omega

\let\p=\partial
\let\<=\langle
\let\>=\rangle

\let\txt=\textstyle

\def\eqn#1{(\ref{#1})}  

\def\beq{\begin{displaymath}}  
\def\eeq{\end{displaymath}}
\def\bea{\begin{eqnarray*}}
\def\eea{\end{eqnarray*}}

\def\ba{\begin{array}}
\def\ea{\end{array}}

\def\o{\over}

\def\comment#1{ \hbox{[{\it Comment suppressed here.}\/]} }
\def\hide#1{}
\def\O{ {\cal O} }

\let\om=\omega

\let\p=\partial
\let\<=\langle
\let\>=\rangle

\let\txt=\textstyle

\def\eqn#1{(\ref{#1})}  

\def\beq{\begin{equation}}
\def\eeq{\end{equation}}
\def\ba{\begin{array}}
\def\bea{\begin{eqnarray}}
\def\ea{\end{array}}
\def\eea{\end{eqnarray}}

\def\comment#1{ \hbox{[{\it Comment suppressed here.}\/]} }
\def\hide#1{}
\def\O{ {\cal O} }

\def\Ord{ {\rm O} }

\def\x{{\bf x}}
\def\y{{\bf y}}
\def\z{{\bf z}}

\def\IR{\relax{\rm I\kern-.18em R}}
\def\IN{\relax{\rm I\kern-.18em N}}
\def\IB{\relax{\rm I\kern-.18em B}}
\def\IE{\relax{\rm I\kern-.18em E}}
\def\ZZ{\relax{\sf Z\kern-.4em Z}}

\def\TT{\mathchoice
       {\sf T\kern-0.52 em{T}}{\sf T\kern-0.52 em{T}}
       {\sf T\kern-0.40 em{T}}{\sf T\kern-0.40 em{T}}}
\def\IP{\mathchoice
       {\sf I\kern-0.14 em{P}}{\sf I\kern-0.14 em{P}}
       {\sf I\kern-0.11 em{P}}{\sf I\kern-0.11 em{P}}}
\def\id{1\kern-.25em {\rm l}}

\def\seq {\! = \!}        

\newcommand{\skipover}[1]{}
\newcommand{\nn}{\nonumber \\}

\def\half {{\txt {1\over 2}}}
\def\third{{\txt {1\over 3}}}

\def\bU {\bar{U}}

\def\psib{{\bar\psi}}

\def\+{\,+\,}
\def\-{\,-\,}

\def\sigF{\si \! \cdot \! F}

\def\SF{Schr\"odinger functional}

\title{
\vskip -94pt
{\small
 \mbox{} \hfill FSU-SCRI-97-117\\
 \mbox{} \hfill September 1997\\
}
\vskip  65pt
       The Schr\"odinger Functional and 
       Non-Perturbative
       Improvement\thanks{Based on talks by R.G.E. and 
       T.R.K.}\thanks{Work supported by DOE grants 
                      DE-FG05-85ER250000 and DE-FG05-96ER40979.}
}

\author{R.G. Edwards${}^{\rm a}$, U.M. Heller${}^{\rm a}$  and
T.R. Klassen\address{SCRI, Florida State University, Tallahassee, FL 32306-4130, USA}
        }

\begin{document}       

\begin{abstract}
After describing the \SF{} for standard and improved gluon and quark actions
we present results for the non-perturbative clover coefficients
of the SW quark action coupled to the Wilson plaquette
action for $\beta \geq 5.7$, as well as the L\"uscher-Weisz one-loop
tadpole improved gauge action, both in the quenched approximation.
\end{abstract}

\kern -10ex
\maketitle

\section{Introduction}\label{sec:intro}

The high cost of lattice QCD simulations has revitalized
interest in the (on-shell) improvement program.
Within the Symanzik~\cite{Sym} 
approach, which we will follow, the use of one-loop
(or even classical) and tadpole~\cite{TI} 
improved gauge actions has lead to
much smaller scaling violations on coarse lattices than for the standard
Wilson plaquette action. 
Numerous studies of the static potential, thermodynamics, 
heavy quarks in either relativistic or non-relativistic frameworks,
and glueballs (the latter on anisotropic lattices) have demonstrated
this. References can be found in the LATTICE proceedings
of the last few years.

The improvement of quark actions is much harder. For
Wilson-type quark actions, which we will consider here, this is
ultimately due to the doubler problem. At least at the quantum
level one incurs $\Ord(a)$ violations of chiral symmetry, which
have turned out to be quite  large.

A great step forward was recently taken by the ALPHA 
collaboration~\cite{ALPHA}, which 
used the \SF{} and the demand that the PCAC relation 
hold at small quark masses, to eliminate {\it all} on-shell $\Ord(a)$ 
errors for Sheikholeslami-Wohlert (SW)~\cite{SW}
 quarks coupled to the Wilson gauge 
action. Various renormalization constants of axial and vector
currents were also calculated non-perturbatively.

The        success of improved gauge actions on coarse lattices
has motivated us to consider the non-perturbative $\Ord(a)$
improvement of quark actions coupled to improved gauge actions. 
In the process we have also reconsidered the case of the SW action
coupled to the Wilson gauge action and extended the determination of 
the O(a) coefficient 
to coarser lattices than in~\cite{ALPHA}.

Although we are also in the process of determining the 
improvement coefficients of various currents, we will
here concentrate on 
the $\Ord(a)$ improvement
coefficient of the {\it action}.
Details of the
general theoretical setup and our motivation can be found in~\cite{TKSF};
our results will be described in detail in future
publications~\cite{EHK}.

\section{$\Ord(a)$ and $\Ord(a^2)$ Improvement}

For Wilson-type quark actions we have to introduce second order
derivative and clover terms to eliminate doublers without introducing 
classical $\Ord(a)$ errors. On the quantum level, on an isotropic lattice,
these two terms are still the only ones that exist at $\Ord(a)$.
We write them as  $a r (\sum_\mu \De_\mu + \half \om \sigF$).
One of the coefficients $r$, $\om$ can be adjusted at will by
a field transformation. It is convenient to fix the Wilson parameter
 $r$; to eliminate all $\Ord(a)$ violations of chiral symmetry
we then have to tune the clover coefficient $\om$ as a function of the
gauge coupling. 

Note that the $\Ord(a)$ terms in the action break chiral but not rotational
symmetry (at this order), whereas the leading $\Ord(a^2)$ errors, that already
exist at the classical level, show the opposite behavior; they break 
rotational but not chiral symmetry.
For this reason the $\Ord(a)$ and leading $\Ord(a^2)$ terms can essentially 
be tuned independently (cf.~\cite{AKL_LAT97}).
By the same token, one can argue that one indeed {\it should} tune
the $\Ord(a)$ and (leading) $\Ord(a^2)$ terms to eliminate the violations
of both chiral and rotational symmetry.  

Eliminating the leading $\Ord(a^2)$ errors in a quark action
leads to the D234 actions~\cite{D234}.   As for gauge actions,
using classical and tadpole improvement at $\Ord(a^2)$ seems to
almost completely eliminate the violation of rotational 
symmetry~\cite{D234}. 

So far we have discussed isotropic lattices. Anisotropic lattices,
with a smaller temporal than spatial lattice spacing, are of great
interest for studies of heavy particles (glueballs, heavy quarks,
hybrids) and thermodynamics. Improvement is more complicated for
actions on such lattices.  
After considering the most general field redefinitions
up to $\Ord(a)$, one sees~\cite{TKSF}
that two more parameters have to be
tuned for on-shell improvement of a quark action up to $\Ord(a)$. One
already appears at $\Ord(a^0)$, namely, a ``bare velocity of light''
that has to be tuned to restore space-time exchange symmetry (by, say,
demanding that the pion have a relativistic dispersion relation for small
masses and momenta). The other is at $\Ord(a)$; the two terms that have
to be tuned at this order can be chosen to be the temporal and spatial
parts of the clover term.

Although the general methods sketched here should eventually be useful
also for the anisotropic case, we will in the following restrict ourselves
to isotropic lattices.

\section{Chiral Symmetry Restoration}

Consider QCD with (at least) two flavors of mass-degenerate quarks.
The idea~\cite{ALPHA}
for determining the clover coefficient is that chiral symmetry
will hold only if its Ward identity is satisfied
as a {\it local operator equation}. In Euclidean space this means that
the PCAC relation between the iso-vector axial
current and the pseudo-scalar density,
\beq\label{PCAC}
  \< \p_\mu A_\mu^b(x) \, \O \> = 2m \, \< P^b(x) \, \O \> 
\eeq
should hold for all operators $\O$, global boundary
conditions, $x$ (as long as $x$ is not in the support of $\O$) 
etc. More precisely, it should hold {\it with the same mass} $m$ 
up to $\Ord(a^2)$ errors (which are quantum errors for (classically and
tapole) $a^2$ improved actions). 
 This will only be the case for the correct value of 
the clover coefficient $\om$.

Several issues have to be addressed before this idea can be implemented
in practice. First of all, even though here we can ignore the multiplicative 
renormalization of $A_\mu^b$ and $P^b$,    
there is an additive correction to $A_\mu^b$ at $\Ord(a)$,
\bea\label{AP}
   P^b(x)     &\, \propto\, & 
                    \psib(x)         \ga_5 \half \tau^b \psi(x) \: , \nn
   A^b_\mu(x) &\, \propto\, & 
                    \psib(x) \ga_\mu \ga_5 \half \tau^b \psi(x)
                     + a \, {c_A} \, \p_\mu P^b(x)         \:.
\eea
The determination of $\om$ is therefore tied in with that of $c_A$. 
We will see later how to handle this.

Note that $\om$ and $c_A$ have an $\Ord(a)$ ambiguity (at least at the
quantum level); different improvement conditions will give somewhat
different values for $\om$ and $c_A$. Instead of assigning an error to
$\om$ and $c_A$ one should choose a specific, ``reasonable'' improvement
condition ---  
the associated  $\Ord(a^2)$ errors in observables are
guaranteed to extrapolate away in the continuum limit.

For various conceptual reasons it is preferrable to impose the PCAC
relation at zero quark mass.
Due to zero modes this is not possible with periodic boundary conditions
(BCs); the quark propagator would diverge.
  Another reason to abandon periodic BCs is that to be sensitive to
the value of $\om$ it would be highly advantageous to have a background
field present; it couples directly to the clover term. The
\SF{} provides a natural setting to implement these
goals. 

\section{The Schr\"odinger Functional} 

The phrase ``Schr\"odinger Functional'' (SF) refers to quantum field
theory with Dirichlet, i.e.~fixed, BCs~\cite{LNWW}.
In the following we will always use periodic BCs in space (extent $L$)
and fixed BCs in time (extent $T$). 
For finite  $T$ the Dirac operator has a gap of order $1/T$ even for 
vanishing quark mass, at least at weak coupling.
 Furthermore, by choosing different BCs at ``opposite
ends of the universe'' one induces a chromo-electric classical
background field.

 In implementing the SF on the lattice the main
point is to understand exactly how to impose fixed BCs on the gauge
and quark fields. In particular, we must be able to do so for
improved actions. For details we refer to~\cite{TKSF}; here we just
mention some salient features:
\vskip 1mm

1. ~The main difference between the Wilson and $\Ord(a^2)$ improved 
gauge actions is that
for the latter the ``boundary'' consists of a double layer of time
slices. To avoid boundary errors larger than those of the bulk action,
one must, already at the classical level, assign loops
at the boundary special ``temporal weight factors'' that depend on the
temporal extent of the loop (cf.~figs.~1 and~2). The classical values of 
the weight factors are easy to understand from elementary calculus formulas
(e.g.~the trapezoidal rule explains the factors of $\half$ in the 
Wilson case of fig.~1). When using the SF as a tool to tune coefficients
in a local action (or current), it is fortunately not necessary to know 
the exact quantum values of the boundary coefficients:
the local Ward identities have to hold 
independent of global effects at the boundary.


\begin{figure}
\vspace*{-4mm} \hspace*{-0cm}
\begin{center}
\epsfxsize = 0.37\textwidth
\leavevmode\epsffile{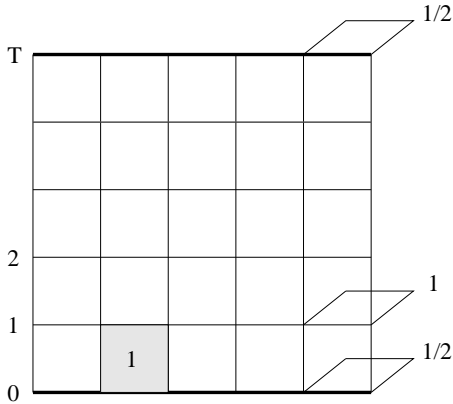}
\end{center}
\vspace*{-1cm}
\caption{Temporal weight factors of various plaquettes in the Wilson
gauge action on a lattice with $T\seq 5a$ and $L\seq 5a$.
}
\end{figure}

\begin{figure}
\vspace*{-2mm} \hspace*{-0cm}
\begin{center}
\epsfxsize = 0.37\textwidth
\leavevmode\epsffile{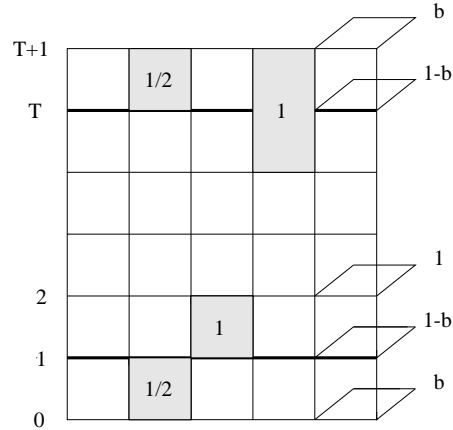}
\end{center}
\vspace*{-1cm}
\caption{As in fig.~1 for an improved gauge action. The thick lines
indicate the inner boundary layers. To avoid classical boundary errors
one must choose $b=1/24$.
}
\end{figure}

2. ~If the boundary values of the gauge field at the top and bottom
of the universe commute, the following is a solution of the lattice field
equations for {\it any} gauge action: $\bU_0(x) = 1$, and
\beq
  \bU_k(x) = \exp\left({1\o T}\left [ x_0 C'_k + (T-x_0) C_k\right]\right) .
\eeq
(The boundary values of the gauge field can be read off from the above
by evaluating it at $x_0=0$ and $T$ for Wilson, respectively, 
$x_0=0,a$ and $T,T+a$ in the improved case.)

The question is if the above background field $\bU$ is the {\it unique}
(up to gauge equivalence) absolute minimum of the classical action for
given boundary values. Uniqueness is important, e.g.~for perturbative
calculations. 
A theorem establishing uniqueness holds in the Wilson case~\cite{LNWW},
if the $C_k$, $C_k'$ parameterizing
the boundary values satisfy certain conditions.
\hide{
are diagonal SU(3) matrices of the form 
$C_k={\rm diag}(\phi_{k1},\phi_{k2},\phi_{k3})/L\, $
(and similarly for $C'_k$) with vectors $\phi_k$, $\phi'_k$ 
in the {\it fundamental domain}, defined by
$\phi_1 + \phi_2 + \phi_3 = 0$, $\,\phi_1 < \phi_2 < \phi_3$, and
$\phi_3 - \phi_1 < 2\pi$.
}
In the improved case it has been checked using simulated annealing that 
uniqueness holds under the same conditions~\cite{TKSF}.

3. ~To impose consistent fixed boundary conditions for the fermion
fields, it is sufficient to consider the projector structure of the
field equations (more precisely, it is only the projector structure
in the {\it time} direction that matters). For an action with the
same projector structure as the standard Wilson quark action, one has
%
to specify $P^+ \psi(x)$, $\psib(x)P^-$ at the (inner) lower
boundary in figs.~1 and~2, and
           $P^- \psi(x)$, $\psib(x)P^+$ at the (inner) upper boundary.
Here $P^\pm \equiv \half (1\pm \ga_0)$. For an improved quark action
with the appropriate projector structure~\cite{TKSF}
one has to specify the same components on both the inner and
outer boundary layers in~fig.~2.

4. ~One of the very useful ideas in applications of the SF is that
of {\it quark boundary fields}~\cite{ALPHA}. They are defined as 
functional derivatives, within the path integral, with respect to
the boundary values specified in the previous paragraph (which are then
set to zero). For the improved case one can actually define two sets of 
quark boundary fields. It turns out that if one defines them with respect
to the {\it outer} boundary values, then most formulas relevant for
our application of the SF are identical for improved and standard
actions. The boundary fields corresponding to the above boundary
values will be denoted as 
$\bar{\zeta}(\x), \zeta(\x), \bar{\zeta}'(\x), \zeta'(\x)$; the first
pair being the {\it lower}, the second the {\it upper} boundary fields.

\section{Details of Non-Perturbative Tuning}

With 
$\O^b =a^6 \sum_{\y\z} \bar\zeta(\y)\, \gamma_5 \frac{1}{2}\tau^b\zeta(\z)$,
in terms of the lower boundary fields, and 
\beq
 f_X(x_0) \equiv -{1\over V} \sum_{b,\x} \, \third\,
      \< X^b(x) \, \O^b \> \:, \, X^b = A^b_0, P^b
\eeq
the PCAC relation becomes
\beq
 m(x_0) \equiv r(x_0) + a\, c_{A} \, s(x_0) = \widetilde{m} + 
 \Ord(a^2) \, .
\eeq
where $ \widetilde{m}$ differs from $m$ by irrelevant multiplicative
renormalization factors, and
\beq
 r(x_0) \equiv {\nabla_0   f_A(x_0)\over 2 f_P(x_0)} \, , \quad
 s(x_0) \equiv {\Delta_0 f_P(x_0)\over 2 f_P(x_0)} \, .
\eeq
Here $\nabla_0$ and $\De_0$ are standard first and second
order lattice derivatives, respectively (in the improved case one actually has
to use an improved first order derivative to be consistent).

Similarly, $f'_X, m', r', s'$ are defined in terms of the upper boundary
fields $\zeta', \bar\zeta'$.

From $\,m(y_0) \stackrel{!}{=} m'(z_0)\,$ one obtains an 
estimator of $c_A$:
\beq
\hat{c}_A(y_0,z_0) ~\equiv~
         -{1\over a} \,\, {r(y_0)- r'(z_0)\over s(y_0) - s'(z_0)} \, .
\eeq
In terms of a suitable $\hat{c}_A$ we now have two different
estimates of the current quark mass:
\bea\label{MMprime}
   M(x_0)  &\,\equiv\, &  r(x_0)  \,+\, a \, \hat{c}_A \, s(x_0) \, , \nn
   M'(x_0) &\,\equiv\, &  r'(x_0) \,+\, a \, \hat{c}_A \, s'(x_0) \, .
\eea
Their equality in the presence of a suitable background field~\cite{ALPHA,EHK}
will be our improvement condition for $\om$.
More precisely, we demand
\beq
  \Delta M(x_0) \equiv M(x_0) - M'(x_0) \stackrel{!}{=} 
                \Delta M^{(0)}(x_0)
\eeq
for some well-chosen $x_0$; here the superscript $(0)$ denotes the 
higher order (and small)
tree-level value of the quantity in question. In practice one measures the
required correlators in a simulation for several trial 
values of $\om$, and  interpolates to find the zero crossing of
$\Delta M - \Delta M^{(0)}$. This determines the non-perturbative
value of $\om$.

\section{Results}

The results we describe in the following all refer to the SW action
on either Wilson or one-loop tadpole improved glue~\cite{LW,Alf}
(which we will refer to as ``LW glue'').
We will always work in the quenched approximation
on isotropic lattices.

Below we use some $\hat{c}_A(y_0,y_0)$ as
estimator of $c_A$ in eq.~\eqn{MMprime}. We then denote 
$\Delta M(x_0)$ by $\Delta M(x_0,y_0)$.
We used lattices $T\cdot L^3=15\cdot 8^3$ or $12\cdot 6^3$ for Wilson
glue, and $14\cdot 8^3$ for LW glue. 
After some study we decided
to use $\Delta M(12,4)$ and $\Delta M(9,3)$, respectively,
in the improvement condition for $\om$ in the Wilson case.
In the improved case we chose  $\Delta M(11,5)$.
Typically we generated $1000-2500$ configurations for each gauge
coupling considered.

Even though the SF alleviates problems due to zero modes, it turns
out that on coarse lattices fluctuations still lead to accidental
zero modes at vanishing quark mass (``exceptional configurations'').
Fortunately, it turns out that 
the mass dependence of the non-perturbative $\om$ is
so weak that one can safely determine it at larger mass values.
In this manner we have extended the non-perturbative clover coefficient
obtained by the ALPHA collaboration for $\beta\geq 6.0$ 
to $\beta \geq 5.7$. In fig.~3 we illustrate the weak mass
(and volume) dependence of $\om$ for $\beta\seq 5.7$ Wilson glue.
We have also checked that the mass dependence is weak for 
$\beta\seq 5.85$ and $6.0$.
 The same can be seen
for LW glue in fig.~4, which furthermore demonstrates the linearity of
$\Delta M$ as a function of $\om$ (all results shown in fig.~4 were 
calculated using the same gauge configurations).

\begin{figure}
\vspace*{-42mm} \hspace*{-0cm}
\begin{center}
\epsfxsize = 0.47\textwidth
\leavevmode\epsffile{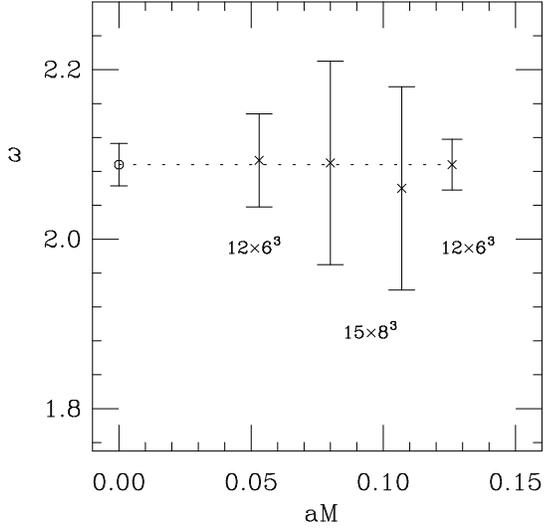}
\end{center}
\vspace*{-1cm}
\caption{The non-perturbative clover coefficient $\om$ for 
 $\beta\seq 5.7$ Wilson glue for different quark masses and volumes,
and its extrapolation to $M\seq 0$.}
\label{Wil_m_w_b570}
\end{figure}

\begin{figure}
\vspace*{-27mm} \hspace*{-0cm}
\begin{center}
\epsfxsize = 0.47\textwidth
\leavevmode\epsffile{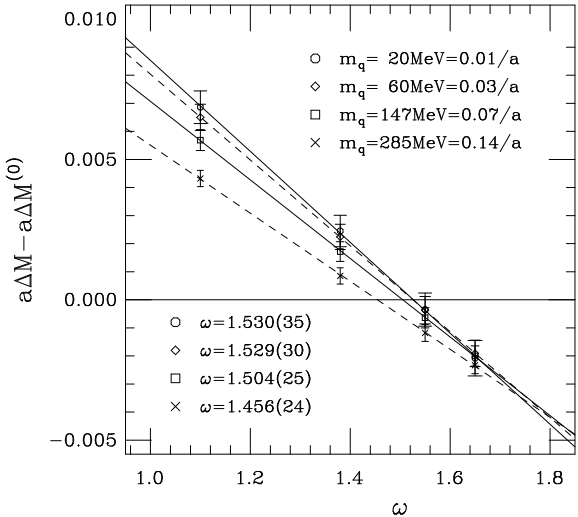}
\end{center}
\vspace*{-2cm}
\caption{
Results from the non-perturbative tuning of $\om$ at different 
quark masses for LW glue at $\beta_{{\rm LW}}\seq 8.4$
(where $a\approx 0.1$~fm).
}
\label{DM_w_fb840}
\end{figure}

For future use of improved quark actions it is advisable to present
the non-perturbative clover coefficient as a definite function of the 
bare gauge coupling. Combining our Wilson results for $\beta=5.7, 5.85$ 
and $6.0$ with those of the ALPHA collaboration and one-loop perturbation
theory we obtain the parameterization ($g^2 = 6/\beta$)
\beq\label{cswWil}
  \om(g^2) = {1-0.6050 g^2 -0.1684 g^4 \o 1 -0.8709 g^2 } \, , \,\,
                                       g^2 < 1.06\, 
\eeq
We were able to accommodate our $\beta=5.7, 5.85$ clover
coefficients and the value for $\beta=6.2$ from~\cite{ALPHA} in our
curve     only   in a 
slightly unsatisfactory manner (extending the Pad\'{e}
in either the numerator or denominator does not help).
This issue is under investigation. In the interim the 
curve~\eqn{cswWil} should be regarded as preliminary.

For the case of LW glue our current data are parameterized well by
\beq\label{cswLW}
  \om_{\rm LW}(g^2) =  {1-0.3590\,g^2 \o 1 -0.4784\,g^2 } \, , \quad 
                                       g^2 < 1.55 \, .
\eeq
(The one-loop coefficient is presently not known analytically
in this case.)
The relation between the coefficient of the plaquette term in the LW action,
$\beta_{\rm LW}$,
and the bare coupling $g^2$ is, to one loop~\cite{LW},
$g^2 = 10/(\beta_{\rm LW} - 1.422)$.
For larger couplings than those in~\eqn{cswLW} it seems that the 
non-perturbative $\om_{\rm LW}$ determined with the SF rises dramatically.
This is currently under investigation.

It is interesting to compare the non\-perturbative clover coefficients
obtained for Wilson and LW glue at the same physical scale. 
The string tension has been measured for both of these actions,
so we will use it to set the scale. We would like a {\it curve}
parameterizing the string tension $\si$ as a function of the coupling.
It is known that the two (or three)
loop running of the coupling does {\it not} properly describe 
$\si$'s
lattice spacing dependence (nor that of other observables)
for the couplings we are interested in. However, as pointed out 
in~\cite{Allton}, this is not to be expected, since $\si$
has 
discretization errors, $g^{2n} a^2, a^4, \ldots$ ($n=0/2$ for Wilson/LW glue)
that should be taken into account. We therefore
try to parameterize $(a \sqrt{\si})(g)$ as
\beq
{\sqrt{\si}\o \La} \, f(g^2) \, \left( \,
  1 + c_2 \, g^{2n} \, \hat{a}(g)^2
    + c_4           \, \hat{a}(g)^4 \, \right),
\eeq
in terms of the three fit parameters $\sqrt{\si}/\La$, $c_2$, $c_4$.
Here $\hat{a}(g)\equiv f(g^2)/f(1)$,
in terms of the universal two-loop function
\beq
a(g)\La = f(g^2) \, \equiv \,
(b_0 g^2)^{-{b_1\o 2b_0^2}} \, \exp(-{1\o 2b_0 g^2})
\eeq

This works very well (for details see~\cite{EHK}), and 
the clover coefficients are presented as
a function of the lattice spacing in fig.~5. 
It is interesting to observe that for $a > 0.05$~fm both the tadpole and
the non-perturbative $\om$'s are pretty much linear in $a$,
at least up to about $0.15$~fm. 
Furthermore, the {\it differences} between the non-perturbative and the
tadpole values are
essentially
linear down to the smallest couplings considered
for both cases (of course, for sufficiently small couplings the
differences should be of order $g^2$).

\begin{figure}
\vspace*{-31mm} \hspace*{-0cm}
\begin{center}
\epsfxsize = 0.47\textwidth
\leavevmode\epsffile{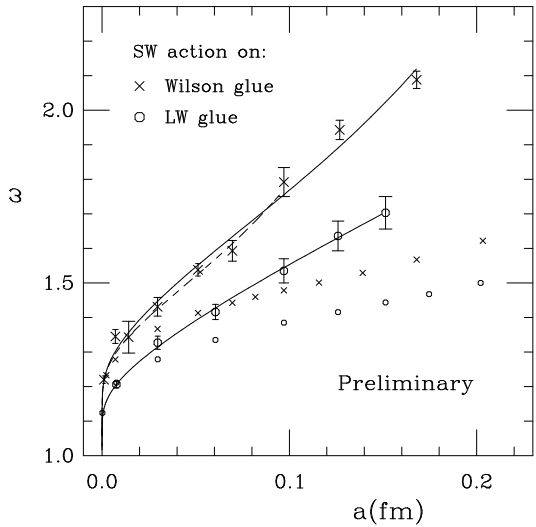}
\end{center}
\vspace*{-1.7cm}
\caption{Non-perturbative clover coefficients versus lattice spacing; 
        scale set by $\sqrt{\si}\seq 460$~MeV. The dashed line denotes 
        the curve from~\cite{ALPHA}. Small symbols indicate
        the tree-level (plaquette) tadpole estimates for $\om$.
}
\label{w_all_noIW_nofb7.75}
\end{figure}

\section{Conclusions and Outlook}

We have shown that the \SF{} and the non-perturbative elimination of
$\Ord(a)$ errors in a Wilson-type quark action can be successfully extended
to improved (gauge) actions.  By establishing that the non-perturbative
clover coefficient has a very weak mass dependence we were able to
determine it for lattice spacings significantly above $0.1$~fm, for
both Wilson and improved gauge actions.

We are currently investigating different definitions of the
axial current improvement coefficient $c_A$. This is
necessary if one wants to determine it on coarse lattices,
where some definitions lead to a $c_A$ of rapidly increasing 
magnitude (at least when using Wilson glue).
Once the $c_A$ determination is completed we plan to determine the
other current normalization and improvement coefficients~\cite{ALPHA}.

There are many other situations in which the non-perturbative elimination
of the $\Ord(a)$ violations of chiral symmetry is important, the most
obvious examples being
full QCD and D234 quarks on anisotropic lattices (for the study of heavy 
quarks).
We hope that the ultimate outcome of this and future studies will be
the ability to perform accurate continuum extrapolations from much
coarser, and therefore cheaper, lattices than hitherto possible.

\end{document}